\documentclass[iop]{emulateapj}
\usepackage{graphicx}
\usepackage{epstopdf}
\usepackage{natbib}
\usepackage{amsmath}

\begin{document}

\title{Present-Day Descendants of $z= 3$ Ly$\alpha$-Emitting Galaxies in the\\
Millennium-II Halo Merger Trees}

\author{Jean P. Walker Soler\altaffilmark{1}, Eric Gawiser\altaffilmark{1}, Nicholas A. Bond\altaffilmark{2}, Nelson Padilla\altaffilmark{3}, Harold Francke\altaffilmark{3}}

\altaffiltext{1}{ Rutgers University, Physics \& Astronomy Department, Piscataway, NJ; jpwalker@physics.rutgers.edu}

\altaffiltext{2}{Goddard Space Flight Center, Astrophysics Science Division, Observational Cosmology Laboratory, Greenbelt, MD}

\altaffiltext{3}{Pontificia Universidad Cat\'{o}lica de Chile, Facultad de F\'{i}sica, Santiago, Chile}

\begin{abstract}
	Using the Millennium-II Simulation dark matter sub-halo merger histories, we created mock catalogs of Lyman Alpha Emitting (LAE) galaxies at $z=3.1$ to study the properties of their descendants. Several models were created by selecting the sub-halos to match the number density and typical dark matter mass determined from observations of these galaxies. We used mass-based and age-based selection criteria to study their effects on descendant populations at $z\simeq 2$, $1$ and $0$. For the models that best represent LAEs at $z=3.1$, the $z=0$ descendants have a median dark matter halo mass of $10^{12.7}$ M$_\odot$, with a wide scatter in masses (50\% between $10^{11.8}$ and $10^{13.7}$ M$_\odot$). Our study differentiated between central and satellite sub-halos and found that $\sim55\%$ of $z=0$ descendants are central sub-halos with $\textup{M}_{\textup{Median}}\sim10^{12}$. This confirms that central $z=0$ descendants of $z=3.1$ LAEs have halo masses typical of $L^*$ type galaxies. The satellite sub-halos reside in group/cluster environments with dark matter masses around $10^{14}\,\textup{M}_\odot$. The median descendant mass is robust to various methods of age determination, but it could vary by a factor of 5 due to current observational uncertainties in the clustering of LAEs used to determine their typical $z=3.1$ dark matter mass.
\end{abstract}

\keywords{galaxies: evolution -- galaxies: formation -- galaxies: high-redshift -- large-scale structure of universe}

\section{Introduction}
	Narrow band surveys have been used to discover Ly$\alpha$-emitting (LAE) galaxies at high redshift (e.g., \citealt{Hu:1996,Cowie:1998,Steidel:2000,Rhoads:2003}) and to study their properties. The galaxies' strong emission lines reveal a set of young, potentially dust-free galaxies theorized by \citet{Partridge:1967}. LAEs allow us to study galaxy formation beginning with one of the smallest building blocks found so far. At $z\simeq 3.1$, typical LAEs are low mass galaxies with $\textup{M}_{Stellar}\simeq10^9\rm \, M_\odot$ and little dust extinction, $A_V\leq0.2$ \citep{Gawiser:2009,Nilsson:2007,Acquaviva:2011a}. These objects have been observed at $z\geq 3$ \citep{Venemans:2005,Gawiser:2006,Gronwall:2007,Nilsson:2007} and as far out as $z\sim 7$ \citep{Iye:2006}.
		
	Previous papers have studied the evolution of LAEs and other, generally more massive, high-z galaxy populations (e.g. \citealt{Gilli:2007,Quadri:2007,Salvadori:2010,Yajima:2011}) by using clustering properties as a technique to make evolutionary claims between redshifts. The connection between $z=3.1$ LAEs and present-day galaxies was determined by \citet{Gawiser:2007} (hereafter Ga07) by measuring the clustering properties of the $z=3.1$ LAEs from the sample of \citet{Gronwall:2007}. Ga07 used the formalism devised by \citet{Mo:1996} to compute the median dark matter mass of their host halos to be $\log_{10}\textup{M}_{\textup{{\tiny DM}}}=10.9^{+0.5}_{-0.9}\,\textup{M}_\odot$. These authors claimed evolution into present-day $L^*$ type galaxies based upon the analytical conditional mass function (e.g. \citealt{Hamana:2006,Francke:2008}), a result echoed for LAEs at $z=2.1$ by \citet{Guaita:2010} and LAEs at $z\simeq6.6$ by \citet{Ouchi:2010}. A weakness of the analytical conditional mass function is its inability to distinguish individual substructures within a halo or to predict the masses of these substructures. N-body simulations fill this gap by producing halo merger trees that allow us to determine properties for the descendant halos and their substructures.
	
	Spectral energy distribution fitting for $z=3.1$ LAEs reveals a large range in stellar ages (Ga07; \citealt{Lai:2008,Nilsson:2007,Ono:2010}). Ga07 determined the young stellar component to have an age of $20^{+30}_{-10}\,\textup{Myr}$ (Ga07) using a two-stellar population model. This could have two simple interpretations: 1) LAEs occur as a galaxy's first burst of star formation, which lasts $\sim 40 \rm \, Myr$, or 2) LAEs are a recurring phase, where each burst of star formation lasts $\sim 40\,\textup{Myr}$ (Ga07). In both cases, stellar evolution produces dust, which ultimately quenches Ly$\alpha$ emission. The old stellar population in the two stellar population model is not well constrained, with an age of up to $2\rm\,Gyr$. A single stellar population model, which observations cannot dismiss, has an age of $50-100\,\textup{Myr}$ (\citealt{Lai:2008}; see also \citealt{Acquaviva:2011a}). \citet{Acquaviva:2011b} determined stellar ages of $\sim1\,\textup{Gyr}$ for $z\simeq3.1$ LAEs. \citet{Nilsson:2007} found $z\simeq3.1$ LAE stellar ages of $0.85^{+0.13}_{-0.42}\,\textup{Gyr}$.
	
	These clustering and spectral energy distribution results allow us to create mock LAE catalogs within the Millennium-II Simulation\footnotemark[1] (\citealt{Boylan:2009}, hereafter MS-II) to study the dark matter mass evolution to the present day within a large cosmological simulation. A similar investigation by \citet{Conroy:2008} used star-forming galaxies at $z\sim2$ as a starting point and studied their dark matter halo evolution until the present-day within the Millennium simulation. Section \ref{MS-II} describes the MS-II. Section \ref{cat} presents specifics of the mock catalogs, while sections \ref{clust} and \ref{descs} report on the clustering analysis and descendants of the catalogs respectively. All distances reported are comoving, $H_0=100h\,\textup{km/s/Mpc}$ and $h=0.73$ is assumed throughout.
\footnotetext[1]{Data created from the MS-II can be accessed from the Max-Planck Institute and Durham University servers at http://www.mpa-garching.mpg.de/galform/millennium-II using a Structured Query Language (SQL) query. For a merger tree explanation see http://www.g-vo.org/Millennium-II/Help?page=mergertrees.}

\section{Simulation and Analysis}
	
\subsection{Millennium-II Simulation}\label{MS-II}
Our study uses the results of the MS-II run by the Virgo Consortium. The MS-II contains $2160^3$ particles in a cube of $100$ $h^{-1}$ Mpc on a side. The particle mass is $6.9\times 10^6h^{-1}\textup{M}_\odot$ with a minimum halo mass of $1.38\times 10^8h^{-1}\textup{M}_\odot$. MS-II gives us the ability to resolve 125 times less massive sub-halos than those observed within the Millennium Simulation \citep{Springel:2005,Lemson:2006}. LAEs at $z=3.1$ appear to be hosted within low mass halos, making the improved mass resolution necessary. Although the Millennium Simulation provides 125 times larger volume, MS-II offers robust statistics for spatial clustering of halos in the mass range of interest. Both simulations use a $\Lambda\textup{CDM}$ cosmology with values:
\begin{eqnarray*}
	&\Omega_{\textup{tot}} =1.0, \Omega_m=0.25, \Omega_b=0.045, \Omega_{\Lambda} =0.75\\
	&h=0.73, \sigma_8 =0.9, n_s=1.
\end{eqnarray*}
All values except $n_s$ and $\sigma_8$ are within $1\sigma$ of the values reported in the 7-year WMAP results \citep{Larson:2011}.

	MS-II also offers improved temporal resolution, with dark matter halos selected in 67 timesteps using a friends-of-friends algorithm \citep{Davis:1985} with a linking length of $b=0.2$ \citep{Boylan:2009}. Each FOF group was analyzed for sub-halos using the SUBFIND algorithm \citep{Springel:2001}, which identifies gravitationally bound sub-halos within the FOF group. We defined a central sub-halo to be the most massive substructure within an FOF group, and other sub-halos within the group were classified as satellites.
	
	During our age selection (described in \textsection \ref{agecat}) we found that some of the youngest dark matter halos exhibited rapid mass growth in the range of a factor of 10-1000 during one timestep (approx. 200 Myr). We believe this mass growth is unrealistic and is caused by misidentification of ownership of dark matter particles by the SUBFIND algorithm between neighboring sub-halos. The errant merger trees appear to normalize by merging with the mass theft victim after a couple of timesteps. This allows us to trust the results from our descendants that have had a few timesteps to normalize. Our selection method, that removes fast growing merger trees, computes the dark matter mass ratios between the $z=3.1$ dark matter halo and its most massive predecessor and between the most massive $z=3.1$ predecessor and its most massive predecessor. If either of these ratios are greater than 10, then we remove this merger tree from our catalogs. With this filtered set of dark matter halos we produce our models.
	
\subsection{Sub-Halo Abundance Matching}
	\label{SHAM}
	We use Sub-Halo Abundance Matching (SHAM) to determine stellar masses for the descendants of our models. The main principle of the SHAM algorithm is to assign luminosities or stellar masses from a luminosity or stellar mass function to dark matter masses from N-body dark matter simulations monotonically \citep[e.g., ][]{Conroy:2006}. Dark matter sub-halos are assigned stellar masses by matching the number densities of halos such that
	\begin{equation}
		\label{eq:principle}
		n_h(>\textup{M}_{\tiny DM})=n_g(>\textup{M}_{\tiny Stellar}(\textup{M}_{\tiny DM})).
	\end{equation}
	The dark matter masses used in eqn. \ref{eq:principle} are sub-halo masses. If the sub-halo is a satellite substructure then we modify the dark matter mass to be the infall mass. Infall mass is defined as the larger of the mass of a central sub-halo before it falls into a larger halo to become a satellite or the satellite mass in the following timestep \citep[See e.g.,][]{Conroy:2006}. We use the infall mass because it gives a better representation of the current stellar component. Unlike the dark matter particles that are disrupted during infall, the stellar component is deeper in the gravitational potential well and therefore is less likely to be disrupted during infall.
	
	Our algorithm solves eqn. \ref{eq:principle} by using a Newton's method approach where the function and its derivative are
	\begin{eqnarray}
		\label{eq:Newton}
		0  & =& f(\textup{M}_{\tiny Stellar}) \nonumber \\
		 & = & n_h(>\textup{M}_{\tiny DM})
		 - \int^\infty_{\frac{\textup{\footnotesize M}_{\tiny Stellar}}{\textup{\footnotesize M}_*}}\phi_*\left(\frac{\textup{M}}{\textup{M}_*}\right)^\alpha e^{-\frac{\textup{\footnotesize M}}{\textup{\footnotesize M}_*}}d\left(\frac{\textup{M}}{\textup{M}_*}\right)  \nonumber  \\ 
		 & = & n_h(>\textup{M}_{\tiny DM})-\phi_* \Gamma\left[1+\alpha,\frac{\textup{\footnotesize M}_{\tiny Stellar}}{\textup{\footnotesize M}_*}\right]  
		 \end{eqnarray}
		 
		 \begin{equation}
		f'(\textup{M}_{\tiny Stellar}) =  \left(\frac{\phi_*}{\textup{M}_*}\right)\left(\frac{\textup{M}_{\tiny Stellar}}{\textup{M}_*}\right)^\alpha e^{-\frac{\textup{\footnotesize M}_{\tiny Stellar}}{\textup{\footnotesize M}_*}}
	\end{equation}
	
	In the above we have used a Schecter function with parameters $\phi_*$, $M_*$ and $\alpha$ and $\Gamma$ is the incomplete gamma function. This method allows us to assign stellar masses by solving eqn. \ref{eq:Newton} for $\textup{M}_{\tiny Stellar}$. The large number of sub-halos in the simulation forces us to use a sub-selection technique to determine the stellar mass to dark matter sub-halo mass function. Afterwards we interpolate this function to determine the stellar masses of the $z=0$ descendants from their sub-halo dark matter masses.

\section{LAE Mock Catalogs}\label{cat}
	We created ten LAE mock catalogs from the MS-II using sub-halos at $z=3.06$ (snapnum\footnotemark[2] 31) based on mass and age selection. The catalog names and properties are listed in Tables \ref{model:mass} and \ref{model:age}. In the following sections we will discuss the selection criteria and motivation for the different catalogs.
	\footnotetext[2]{Snapnum is the snapshot (timestep) number from the MS-II.}

\subsection{Mass-based Catalogs}\label{masscat}
	All mass-based catalogs were chosen to reproduce the $z=3.1$ LAE number density ($1.5\times 10^{-3}\textup{Mpc}^{-3}$) determined by \citet{Gronwall:2007}, thus generating models with 3856 sub-halos in the MS-II volume.
	\begin{enumerate}
		\item The mass limit criterion was selected to have a minimum mass of $\log_{10}\textup{M}_{\textup{\tiny Min}}=10.6\,\textup{M}_\odot$ to match that inferred by Ga07 from the observed clustering of $z= 3.1$ LAEs. This approach is similar to the one used by \citet{Conroy:2008}; their mass limit was chosen to reproduce the clustering of $z\sim2$ star-forming galaxies. We randomly selected 5.24\% of these sub-halos to match the above number density. The Mass limit catalog was designed to reproduce the observed correlation length of $z= 3.1$ LAEs.
		\item The Median catalog was selected using the median mass reported by Ga07, $\log_{10}\textup{M}_{\textup{\tiny Med}}=10.9\,\textup{M}_\odot$, as the catalog's median mass value. We expanded evenly around the median mass to obtain the observed LAE number density, choosing halos with a range of $7.56\times 10^{10}\,\rm M_\odot$--$8.36\times 10^{10}\,\rm M_\odot$. As expected, this catalog and the Mass Limit catalog have similar median masses; it should also reproduce the observed correlation length.
		\item The $-\sigma$ and $+\sigma$ catalogs were selected and named based on the $\pm1\sigma$ uncertainty reported by Ga07 in the observed LAE median mass. The median masses of the catalogs are $\log_{10}\textup{M}_\textup{\tiny Med}=10.0\,\textup{M}_\odot$ and $11.4\,\textup{M}_\odot$, respectively. We expand evenly around their respective median masses to obtain the observed LAE number density. The mass ranges for the $-\sigma$ and $+\sigma$ catalogs are $9.93\times 10^{9}\,\rm M_\odot$--$1.01\times 10^{10}\,\rm M_\odot$ and $2.15\times 10^{11}\,\rm M_\odot$--$3.04\times 10^{11}\,\rm M_\odot$, respectively. These catalogs were created to study the uncertainty in the descendant properties propagated from the observed clustering uncertainties and are not expected to reproduce the best-fit observed correlation length at $z=3.1$.
	\end{enumerate}

\subsection{Age-based Catalogs}\label{agecat}
\begin{figure}[h]
	\begin{center}
		\includegraphics[scale=.65]{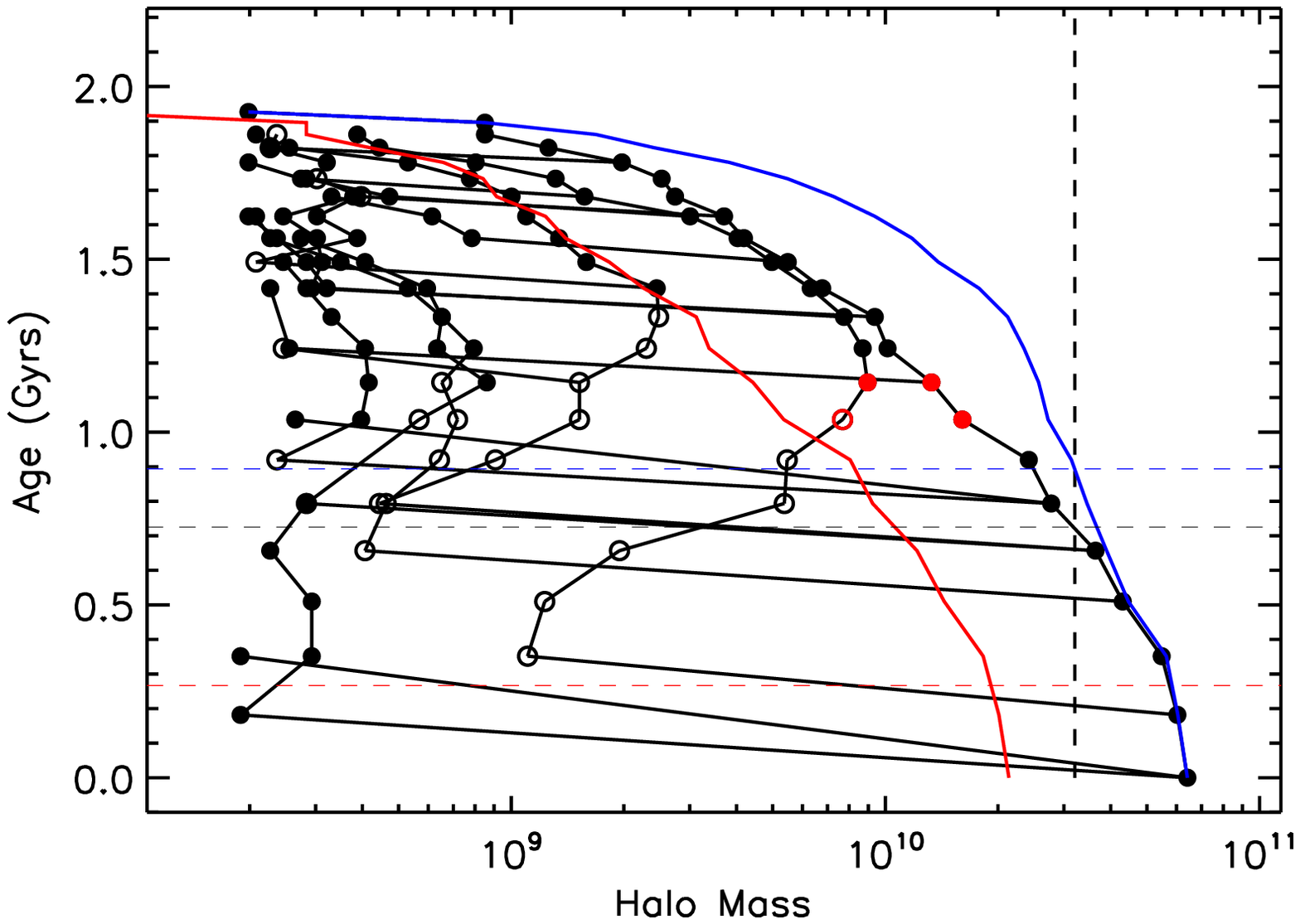}
		\caption{Representation of a merger tree which shows the ages found for one particular $z=3.1$ central sub-halo. Sub-halos are represented by circles, where filled circles are centrals and open circles are satellites. The black, blue and red horizontal dashed lines mark the Formation, Assembly and Merger ages respectively. The black vertical dashed line marks half the maximum mass of sub-halos in the merger tree. The Formation age is assigned when the most massive sub-halo crosses this vertical line. The blue solid curve is the sum of the sub-halo masses at each timestep and the Assembly age is marked by the intersection of this curve with the black vertical line. The red solid curve marks the left boundary of the region where a 3:1 ratio between the most massive object and any other sub-halo can occur. The sub-halos involved in the beginning of the $\leq 3:1$ major merger are marked with red circles. The Merger age is assigned when the $\leq 3:1$ ratio satellite disappears due to having fully merged with the central.}
		\label{tree2}
	\end{center}
\end{figure}

\begin{figure}[h]
	\begin{center}
		\includegraphics[scale=.9]{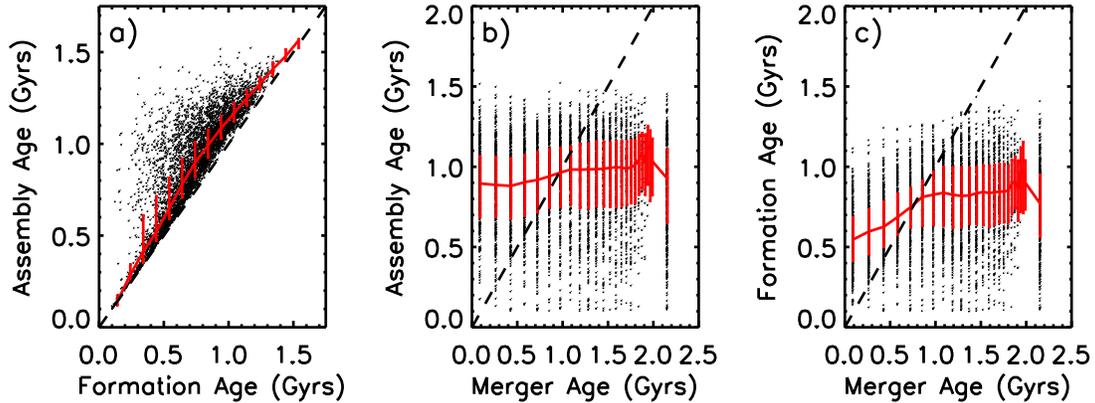}
		\caption{Figures show the relationships between the different age definitions. Red curves are the median with 25th and 75th percentiles shown as red error bars. Panel a: Shows the relation between the Assembly and Formation ages. By definition the Formation age is equal to or greater than the Assembly age. Panel b: Shows the relation between the Assembly and Merger ages. There is no correlation between these two age definitions. Panel c: Shows the relation between Formation and Merger ages. There is little to no correlation between these two age definitions. The merger ages are discretized due to averaging of timesteps (See $\mathsection$\ref{agecat}).}
		\label{agevage}
	\end{center}
\end{figure}

	Three age definitions were chosen to study the dependence of descendants' properties on age selection. The age definitions use merger trees rooted in central sub-halos within FOF halos with $\textup{M}_\textup{{\tiny FOF}}\ge 3.98\times10^{10}\,\textup{M}_\odot$, the mass limit criteria. Figure \ref{tree2} shows an example merger tree showing the formation, assembly and merger ages assigned. Ages are defined as the difference of lookback time chosen and the lookback time at $z=3.1$. The three age definitions are as follows:
\begin{enumerate}
	\item The {\it Formation age} \citep{Gao:2004,Gao:2005} quantifies the timescale for mass growth of the most massive dark matter structure. It is assigned by finding the most recent timestep where the sub-halo mass of the most massive progenitor is less than half of the maximum mass in the entire merger tree. If this occurs between two timesteps, we linearly interpolate between these two to estimate the time when half the mass was accreted.
	\item The {\it Assembly age} \citep{Navarro:1997} measures when half the maximum mass of a galaxy is present in collapsed sub-halos even if they have not yet merged. This is assigned by finding the most recent timestep where the sum of progenitor sub-halo masses is less than half of the maximum mass in the entire merger tree. If this occurs between two timesteps, we linearly interpolate between these two to estimate the time when half the mass has assembled. Because of these definitions, the Assembly ages are always equal to or greater than the Formation ages. Figure \ref{agevage} a) shows the relation between Formation and Assembly ages.
	\item {\it Merger age} searches for the most recent major merger. Our definition was designed to to use a main sub-halo to follow the major merger which is similar to the methods used by \citet{Genel:2008}. We define a  major merger to occur when two central sub-halos have a mass ratio of 3:1 or less in a timestep and the most massive halo within that timestep is involved. The major merger begins when one of the descendants in the next timestep is a satellite of the other descendant or both centrals merge into a single sub-halo. In Figure \ref{tree2} we have two central sub-halos, around $1.2\,\textup{Gyr}$, where the lower mass central sub-halo descends into a satellite belonging to the descendant of the more massive central sub-halo. We track the descendants until the two sub-halos merge, possibly triggering star formation. We average the timestep of this sub-halo merger and the previous timestep to assign the merger time. The infall timescale from the beginning of the 3:1 merger of the two centrals to the final merger of their descendants is comparable to the dynamical friction timescale used by \citet{Genel:2008} based on work from \cite{Boylan:2008}. Figure \ref{agevage} b) \& c) show little to no correlation between Merger age and Assembly or Formation ages. 
\end{enumerate}

\begin{figure}[h]
	\begin{center}
		\includegraphics[scale=0.95]{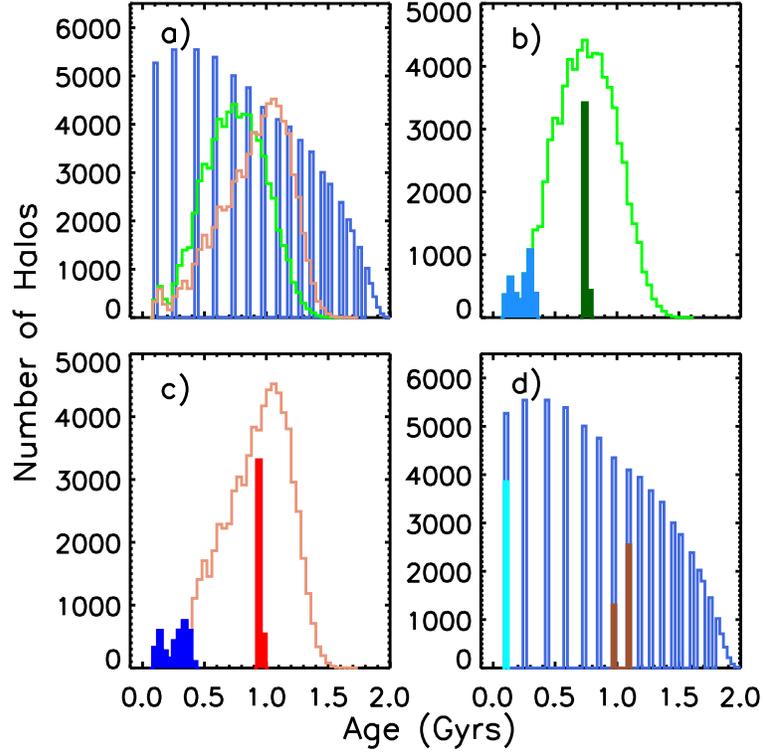}
		\caption{Light green, peach and royal blue curves are the age distributions for central sub-halos at $z=3.1$ with M$_{\textup{\tiny FOF}}\geq 3.98\times 10^{10}$ M$_\odot$ using the Formation, Assembly and Merger definitions. Panel a: Shows the three age distributions. Panels b,c,d show the Formation, Assembly and Merger age distributions (same colors as panel a) along with those of the selected young and median aged mock halo models in different colors.}
		\label{agedist}
	\end{center}
\end{figure}

	The age distributions are presented in Figure \ref{agedist} a). The LAE mock catalogs are then created from the ages calculated for all halos with the mass limit criteria by selecting the 5.24\% youngest and 5.24\% closest to median aged sub-halos. See Figure \ref{agedist} b), c) \& d) for age distributions for all halos and the age distributions of the selected catalogs. Table \ref{model:age} lists the age properties of the different catalogs with the bold values being age statistics used to select that particular catalog.
	
	All the median age-selected catalogs have median ages of $\sim 1 \,\rm Gyr$. For star formation triggered by accretion of sub-halos in minor mergers, Formation and Assembly ages should roughly track the age of stars in a galaxy i.e. the population dominating its stellar mass \citep[see e.g.,][]{delucia:2007}. Since SED fits allow $\sim1\,\textup{Gyr}$ old populations to comprise the majority of stellar mass in LAEs (Ga07; \citealt{Acquaviva:2011b,Nilsson:2007}), the median Formation and Assembly models are feasible models. However, because the merger age should track a young stellar population born in a starburst triggered by the major merger, we cannot reconcile a $\sim1\,\textup{Gyr}$ median merger age with SED results; this rules out the Median Merger model at high confidence. The young Formation and Assembly catalogs are viable models which have median ages of 240 Myr, consistent with ages found for single-population models. The young Merger catalog has a median age of 91 Myr and is consistent with the observed age of the young stellar component.
		
\section{Clustering Analysis of Mock Catalogs}\label{clust}
We used the naive estimator $\xi_N$ (e.g., \citealt{Landy:1993}) to calculate the two point auto-correlation function (2PCF) for the models.
\begin{equation*}
	\hat{\xi}_{N}(r)=\frac{DD(r)}{RR(r)}-1 
\end{equation*}
\begin{eqnarray}
	DD(r)&=&\frac{2\times\textup{\footnotesize Number of data-data pairs within a radial bin}}{n_D(n_D-1)}\nonumber \\
	RR(r)&=&\frac{\textup{\footnotesize Volume contained within a radial bin}}{\textup{\footnotesize Volume of simulation}} \nonumber
\end{eqnarray}
Data-data pairs are unique pairs of sub-halos, from a catalog, that are separated and binned by radius $r$. The simple geometry of the simulation, in conjunction with the periodic boundary conditions, allow us to use a geometrical formula for RR, which eliminates uncertainty in the estimator caused by binning random-random pairs. 

	We applied a correction to our data abscissas to match the effective center of the radial bin using 
\begin{eqnarray}
	\label{radialcorrection}
	<\xi>_{bin}=\frac{\int_{r_L}^{r_L+\Delta r} \left(\frac{r}{r_0}\right)^{-\gamma}r^2dr}{\int_{r_L}^{r_L+\Delta r}r^2 dr} \\ \nonumber\\
	r_{bin}=r_0<\xi>_{bin}^{-1/\gamma}.
\end{eqnarray}
The correction was applied by our fitting algorithm to shift the data point to the radius which corresponds to the average value of the bin. We fit our naive estimator with the power law given by
\begin{eqnarray}
	\label{model:eq}
	\xi (r)=\frac{(r/r_0)^{-\gamma}-\omega_{\Omega}}{1+\omega_\Omega} \\
	\omega_{\Omega}=\int^{R_{max}}_0RR(r)\left( \frac{r}{r_0}\right)^{-\gamma} dr,
	\label{intcon}
\end{eqnarray}
where $\omega_{\Omega}$ is the integral constraint found from the power-law term, $(\frac{r}{r_0})^{-\gamma}$ for the parameters during fitting. We minimized $\chi^2$ to determine the best fit for the parameters $r_0$ and $\gamma$ using 
\begin{eqnarray}
	\label{error}
	\sigma_{N}^2 (r) & = & \left(\frac{1+\xi(r)}{1+\omega_\Omega}\right)^2\left(\frac{1-RR(r)}{(n_D(n_D-1)/2)RR(r)}\right)
\end{eqnarray}
for the variance of the naive estimator \citep{Landy:1993}. 

\begin{figure}[h]
	\begin{center}
		\includegraphics[scale=.75]{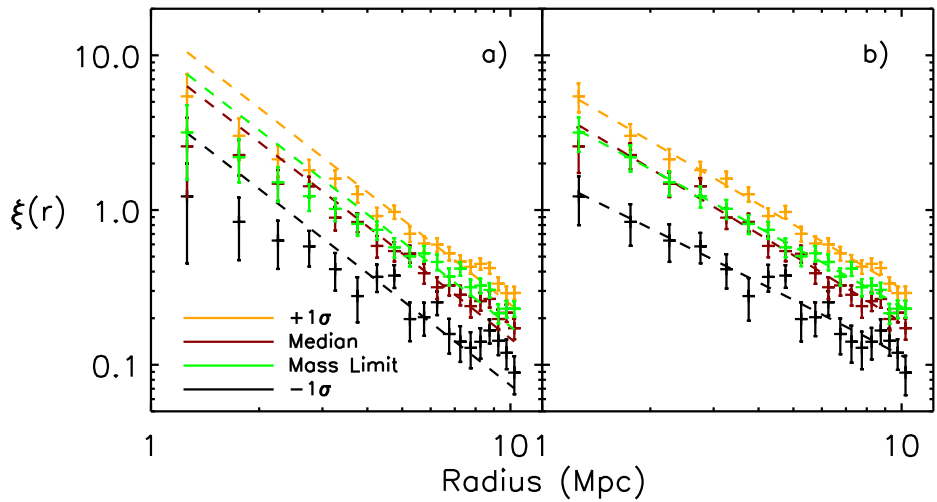}
		\caption{The two-point autocorrelation functions for mass-based mock models. See Table \ref{model:mass} for correlation length and gamma values. Error bars are a function of the model being fitted. Panel a): We fitted a power-law with $\gamma=1.8$ to the data to determine the correlation length for each model. As expected, the models show a trend to larger correlation length with higher median mass. Panel b): We fit a general power-law to determine the best fit correlation length and  $\gamma$. We find $\gamma\simeq 1.4$ for these mock LAE models.}
		\label{masscorr}
	\end{center}
\end{figure}

\begin{figure}[h]
	\begin{center}
		\includegraphics[scale=.85]{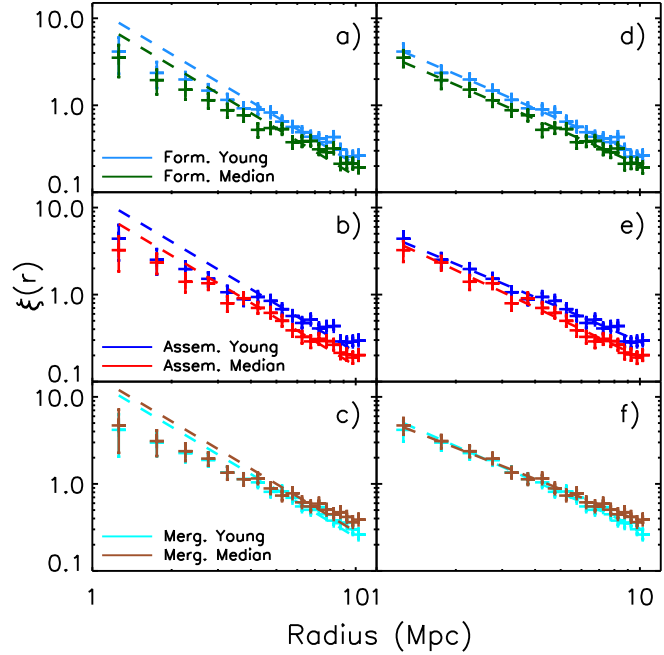}
		\caption{Two-point autocorrelation functions for age-based catalogs. See Table \ref{model:mass} for correlation length and gamma values. Panels a, b, and c): We fit a power-law with $\gamma=1.8$ to the data to determine the correlation length for each model. The correlation length of the formation and assembly age models separate based on age selection. The merger age definition shows a smaller separation based on age selection. Panels d, e, and f): We fit a general power-law to determine the best fit correlation length and  $\gamma$. We find $\gamma\simeq 1.4$ for these LAE catalogs. We find an age-clustering relation when we use the formation and assembly ages where the youngest dark matter halos are more clustered than the median aged dark matter halos.}
		\label{agecorr}
	\end{center}
\end{figure}

	In the following discussion we will fix $\gamma=1.8$ to compare our results to the observed correlation length for $z=3.1$ LAEs, $r_0=3.6^{+0.8}_{-1.0}\,\textup{Mpc}$ (Ga07). Table \ref{model:mass} lists the values found for $r_0$ fixing $\gamma=1.8$ and also fitting both $r_0$ and $\gamma$ as free parameters, while Figures \ref{masscorr} and \ref{agecorr} show the best fit 2PCF for the catalogs.

\subsection{Clustering of Mass-Based Catalogs}
Figure \ref{masscorr} and Table \ref{model:mass} show the mass-based catalogs' 2PCFs, correlation lengths, $\gamma$ and $\chi^2$ values. We confirm that correlation lengths increase with median mass in our mass-based catalogs for both fixed $\gamma=1.8$ and when $\gamma$ is allowed to be a free parameter. The Median and Mass limit catalogs have similar median masses and were expected to have the same correlation lengths when $\gamma =1.8$; the correlation lengths are consistent with one another. These two models also have correlation lengths that are consistent with the observed LAE correlation length of $r_0=3.6^{+0.8}_{-1.0}\,\textup{Mpc}$, making them good representations of LAEs at $z=3.1$. The $+\sigma$ and$-\sigma$ models have correlation lengths close to the $\pm1\sigma$ uncertainty from observed values, as expected. Fixing $\gamma=1.8$ allows us to compare our results to previous works, but the best fit models are not consistent with this value. We find an average value of $\gamma=1.33$, which is consistent with the results from \citet{Boylan:2009} at $z=2.07$ using our fitting range.

\subsection{Clustering of Age-Based Catalogs}
The similarities of the mass properties of the age-based catalogs (see Table \ref{model:mass}) imply that any difference in the clustering properties between age-based catalogs are due to the age definition and selection. The Formation Median and Assembly Median models have similar correlation lengths of $r_0\simeq3.6\,\textup{Mpc}$ and are consistent with the observed LAE correlation length. The Assembly Young and Formation Young have higher correlation lengths of $r_0\simeq 4.4\,\textup{Mpc}$  and are barely consistent at the 68\% level with the $z\simeq3.1$ LAE correlation length. We cannot rule out the Young Merger model at 95\% confidence. The Median Merger model has a correlation length large enough to be ruled out at 93\% confidence. This large correlation length and the high merger ages make the Median Merger model an unrealistic $z\simeq3.1$ LAE model. No other models are ruled out based on clustering. The observed trend for the Formation and Assembly ages that young aged models have a larger clustering length compared to the median aged models does not disagree with the findings of \citet{Gao:2005} that the youngest $20\%$ of halos at $z=0$ cluster less strongly than the oldest $20\%$. Once we use a similar selection we also find that the oldest halos have higher clustering compared to the youngest.

\section{Descendants of Mock LAEs}\label{descs}
	We used the MS-II merger trees to find the $z=2.07$ (snapnum 36), $z=0.99$ (snapnum 45), and $z=0$ (snapnum 67) descendants of the mock LAE catalogs. We report the mass distributions of the descendants based on the mass of the host FOF group, $\rm M_{\textup{\tiny FOF}}$. We prefer M$_{\textup{\tiny FOF}}$ because it determines the evolution of dark matter halos and is used in the Press-Schechter formalism, while the individual sub-halo mass traces the evolution of individual galaxies. We classify the most massive sub-halo within each FOF group as a central and other sub-halos as satellites. All other smaller sub-halos within the FOF group are satellites. The following sub-sections will discuss the descendants of our mock catalogs as reported in Tables \ref{model:massdesc2}-\ref{model:massdesc0}.

\subsection{Mass Limit and Median Mass Model Descendants}\label{med}
\begin{figure}[h]
	\begin{center}
		\includegraphics[scale=.65]{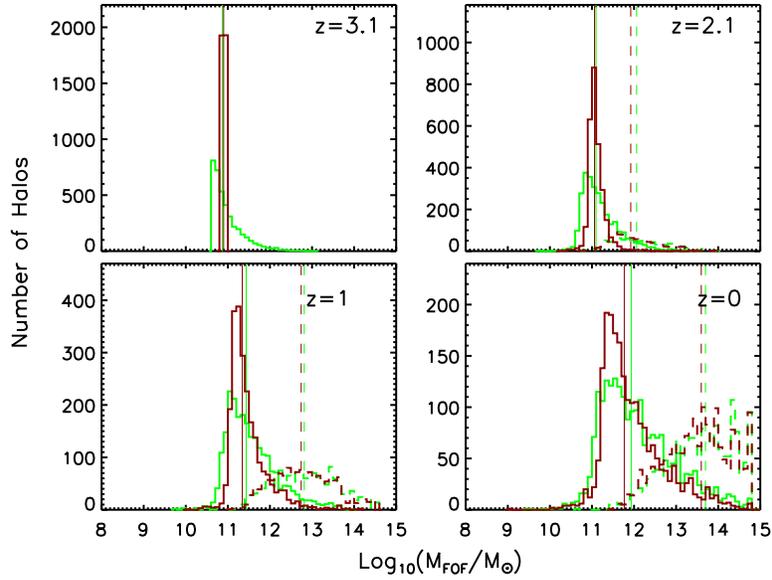}
		\caption{Mass histogram, using FOF mass, for the Median (dark red) and Mass limit (green) mock catalogs at $z=3.1$ (upper-left) and descendants at $z=2.1$ (upper-right), $z=1$ (lower-left) and $z=0$ (lower-right). Solid and dashed curves represent the central and satellite mass distribution breakdown. Satellite sub-halos have a larger mass than central sub-halos because we plot M$_{\textup{{\footnotesize FOF}}}$, the mass of the FOF group to which the sub-halo belongs. Vertical dashed lines delineate the median mass of each distribution.}
		\label{masslimitmedianfof}
	\end{center}
\end{figure}

\begin{figure}[h]
	\begin{center}	
		\includegraphics[scale=.95]{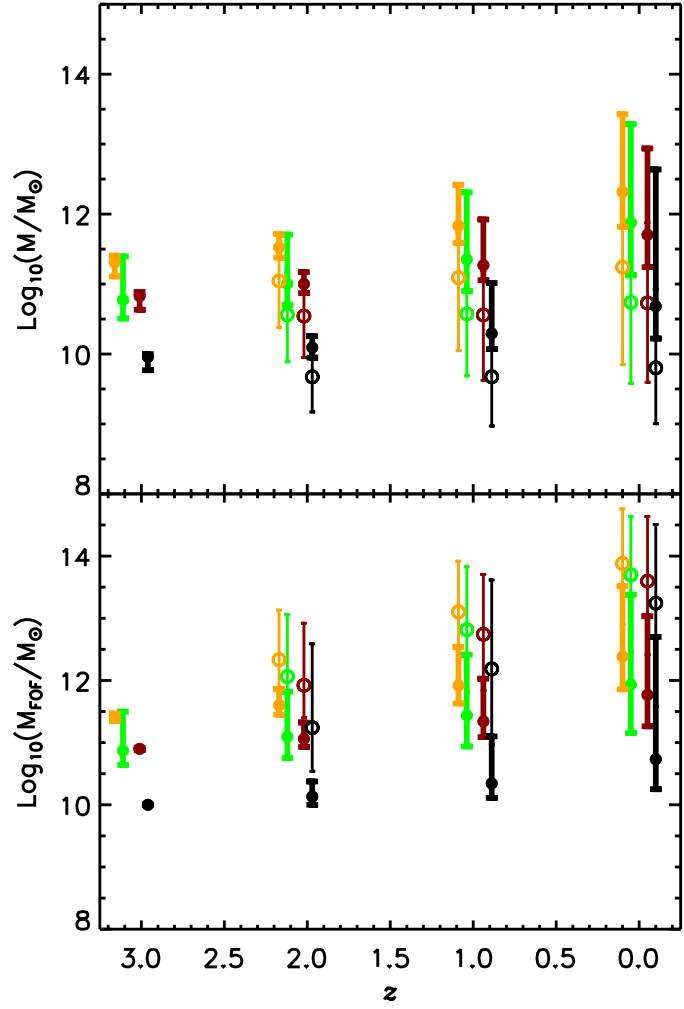}
		\caption{Figure shows the sub-halo (top panel) and FOF (bottom panel) mass evolution of mass selected models through redshift. The Median (dark red), Mass limit (green), $+1\sigma$ (orange) and $-1\sigma$ (black) models are evenly spaced around their redshift for easier viewing. Solid and open points represent the distributions for centrals and satellites, respectively. Points are the median sub-halo masses while the error bars show the 10th and 90th percentiles for the distributions. The sub-halo and FOF masses for centrals tend to increase towards lower redshift as sub-halos merge and accrete. Some of the $z=3.1$ central halos merge with more massive halos becoming satellites. These satellites reside in massive FOF halos, but have a smaller individual sub-halo mass.}
		\label{massplot}
	\end{center}
\end{figure}

	Figure \ref{masslimitmedianfof} shows an example histograms demonstrating the evolution of the Mass Limit and Median models from $z=3.1$ to $z=0$ using FOF halo masses. Both models have the same median masses, $\log_{10} \textup{M}_{med}\simeq 10.9\,\textup{M}_\odot$, at $z=3.1$, as also seen in Table \ref{model:mass}. All other models evolve in a similar manner. Figure \ref{massplot} (bottom panel) summarizes   this same evolution for all of the mass selected models. The models' descendants also have similar satellite mass distributions (dashed histograms in figure \ref{masslimitmedianfof}) as they evolve with time. The satellite median masses grow from $\textup{M}_{\textup{{\tiny FOF; Med}}}=10^{12}\textup{M}_\odot$ at $z=2$ to $10^{13.7}\textup{M}_\odot$ at $z=0$. The central descendants show evolution towards higher masses, as expected for bottom-up halo growth \citep{Davis:1985}, though the 10th percentile in mass stays roughly constant at its $z\simeq 3.1$ value. We find that the catalogs' central population, which comprise 55-60\% of the descendants, has mass growth from $\textup{M}_{\textup{{\tiny FOF; Med}}}=10^{10.9}\,\textup{M}_\odot$ at $z= 3.1$ to Milky Way-sized dark matter halos, $\simeq 10^{11.8}\textup{M}_\odot$, at $z=0$. The median FOF mass for all descendants grows from $\textup{M}_{\textup{{\tiny FOF; Med}}}=10^{10.9}\,\textup{M}_\odot$ at $z= 3.1$ to $\textup{M}_{\textup{{\tiny FOF; Med}}}=10^{12.6}\,\textup{M}_\odot$ at $z=0$.
		
	For comparison, Figure \ref{massplot} top panel shows these descendants in terms of the individual sub-halo mass instead of the mass of the FOF group that the sub-halo resides in. The evolution of mass in this sense is similar to the FOF mass evolution; sub-halos tend to grow in mass towards $z=0$, but mass loss is seen in the satellite populations. Central sub-halo masses are similar to the FOF masses; therefore, we expect a similar median mass at $z=0$. We find the central sub-halo median mass to be $\textup{M}=10^{11.8}\textup{M}_\odot$ at $z=0$. Comparing the top and bottom panels of figure \ref{massplot} we see that descendants that are satellites reside in massive halos, but the satellite sub-halos are less massive, with a median mass of $10^{10.8}\,\rm M_\odot$ at $z=0$. 
	
\subsection{Age-Based Model Descendants}
\begin{figure}[h]
	\begin{center}
		\includegraphics[scale=.95]{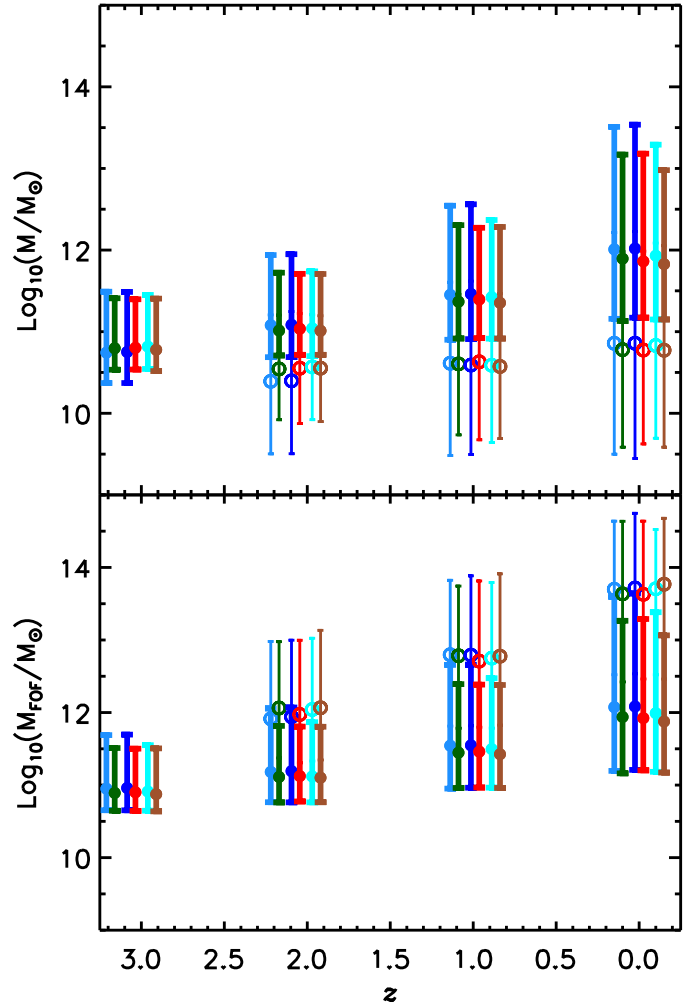}
		\caption{Figure shows the sub-halo (top panel) and FOF (bottom panel) mass evolution for our age selected models. All the age selected models have similar $z=3.1$ masses. Median sub-halo mass values tend to increase towards low redshift as the dark matter halos merge and accrete. There is a small difference in the median masses between the young formation/assembly and median formation/assembly due to age selection. This small mass difference propagates to $z=0$ in the median masses of the centrals. Otherwise these models have nearly the same distribution at each redshift. Some of the $z=3.1$ central halos merge into more massive FOF halos producing satellites at lower redshift. These satellites have smaller sub-halo masses due to merging into a more massive FOF halo.}
		\label{massplotage}
	\end{center}
\end{figure}

	The similarities in the six age-based models' mass distributions at $z=3.1$ suggests that we should expect similar descendant distributions. Figure \ref{massplotage} top and bottom panel confirm the similarities among these catalogs' descendants. The mass evolution represented in figure \ref{massplotage} bottom panel is the mass of the FOF halo where the sub-halo resides while the top panel uses sub-structure mass. The evolution of the mass distributions are all similar and therefore independent of age definition or selection. The descendant central median masses are $\textup{M}_{\textup{{\tiny FOF}}}=10^{11.2}\,\textup{M}_\odot$ at $z=2.1$, $10^{11.5}\,\textup{M}_\odot$ at $z=1$, and $10^{12}\,\textup{M}_\odot$ at $z=0$ with a small scatter of less than a factor two. We find that the fraction of central sub-halos for all age-based models is independent of age definition or selection (Tables \ref{model:massdesc2}-\ref{model:massdesc0}). The satellite descendants at $z=0$ comprise $42-44\%$ of the population and have a median FOF mass of $10^{13.7}\textup{M}_\odot$. For all the models the full distribution of descendants have a median FOF mass within a factor of two of $10^{12.7}\textup{M}_\odot$ at $z=0$. These FOF masses are in agreement with those found for the Median and Mass limit catalogs described in section \ref{med}. The sub-halo masses of the age-based models are also similar to the values obtained for the Median and Mass limit models.

\subsection{$+\sigma$ and $-\sigma$ Model Descendants}
	The $+\sigma$ and $-\sigma$ models were created to quantify the uncertainties within the descendant distributions caused by the uncertainties in the observed clustering analysis. Figure \ref{massplot} shows the $+\sigma$ and $-\sigma$ with the other mass-selected models for comparison using the FOF mass. The central descendant sub-halos have median masses that are a factor of four higher for the $+\sigma$ model and a factor of ten lower for the $-\sigma$ model compared with the Median mass catalog's descendants at $z=0$. The satellite distributions at $z=0$ have median masses that are a factor of three larger for the $+\sigma$ and a factor of five smaller for the $-\sigma$ than the Median model. When we consider the full descendant distribution the median mass of the $+\sigma$ model is a factor of three larger and the $-\sigma$ model is six times lower than the Median model. The fraction of central descendant sub-halos is similar between the two models.
	
\section{Determining Stellar Properties of Descendants}
\label{schechter}
\begin{figure}[h]
	\begin{center}
		\includegraphics[scale=0.72]{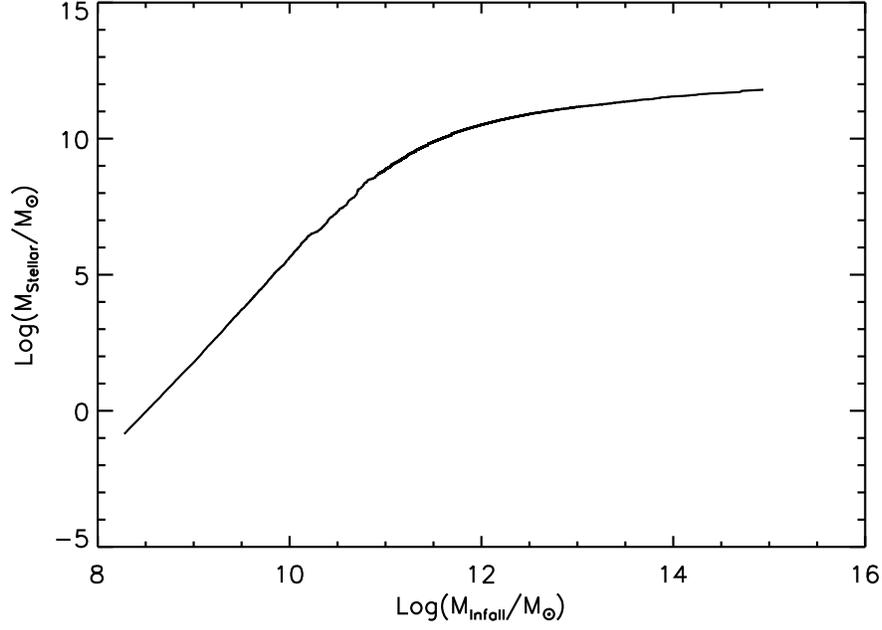}
		\caption{Relation between infall and stellar mass found from the SHAM algorithm after application to all $z=0$ sub-halos. We use this relation to determine the stellar masses from the $z=0$ descendant sub-halos. The stellar mass function used in the SHAM algorithm had a fitting range of $10^{8.5}-10^{11.85}\,\textup{M}_\odot$ \citep{panter:2007} and stellar masses below and above these values are extrapolations from the SDSS fit.}
		\label{infallstellar}
	\end{center}
\end{figure}

Using the SHAM algorithm discussed in section \ref{SHAM}, we determine the stellar masses for all sub-halos at $z=0$. We use the Schecter function parameters found by the Sloan Digital Sky Survey \citep{panter:2007}. The parameters are
\begin{eqnarray*}
	\phi^* & = & 2.2\pm0.5_{stat}\pm1_{sys}\times 10^{-3}\,\textup{Mpc}^{-3} \\
	\textup{M}^* & = & 1.005\pm0.004_{stat}\pm0.200_{sys}\times 10^{11}\,\textup{M}_\odot \\
	\alpha & = & -1.222\pm0.002_{stat}\pm0.1_{sys}.
\end{eqnarray*}
Figure \ref{infallstellar} shows the relationship between infall and stellar masses determined from the SHAM algorithm.

\begin{figure}[h]
	\begin{center}
		\includegraphics[scale=0.65]{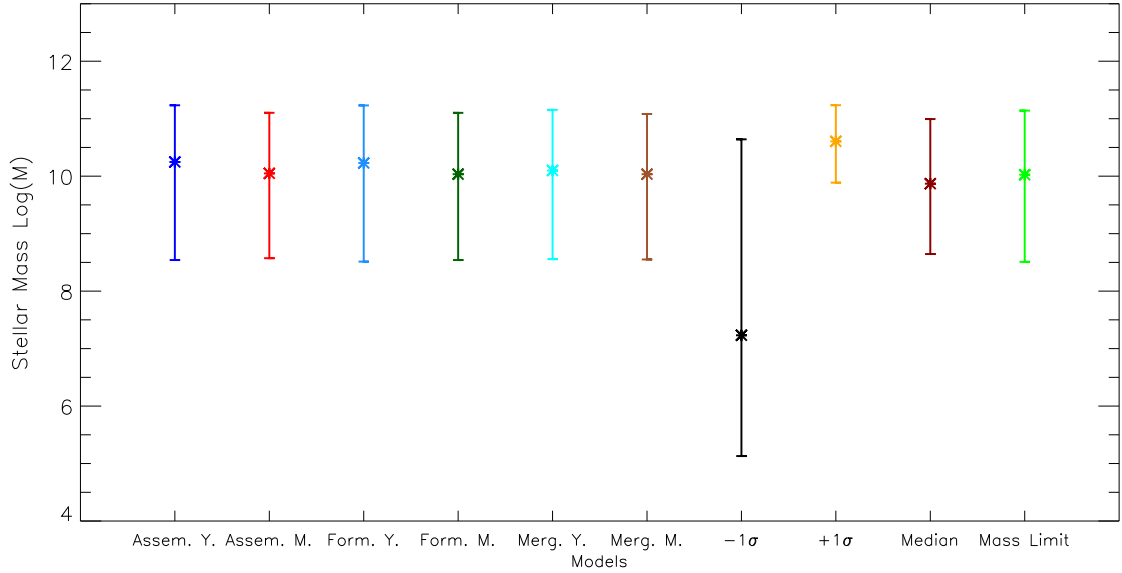}
		\caption{Median stellar masses of the descendants at $z=0$. The error bars denote the 10th and 90th percentile stellar masses for each of the models' distributions. All models except the $+1\sigma$ and $-1\sigma$ models have similar stellar mass distributions. The low stellar masses found in the $-1\sigma$ model are caused by the extrapolation of the stellar mass function towards low mass (See $\mathsection$\ref{schechter}).}
		\label{stellarmass}
	\end{center}
\end{figure}

	After application of the SHAM algorithm to the entire $z=0$ MSII sub-halo catalog covering dark matter masses ranging from $10^{8.28}\,\textup{M}_\odot$ to $10^{14.94}\,\textup{M}_\odot$, we obtained stellar masses ranging from $0.14\,\textup{M}_\odot$ to $10^{11.80}\,\textup{M}_\odot$. The very low stellar masses are attributable to the excess of dark matter sub-structures at low mass compared to the number of galaxies from the stellar mass function, which remains an unsolved problem in galaxy formation. The $-1\sigma$ model is the only model with a significant number of halos in this range. Figure \ref{stellarmass} shows the median stellar masses for all the models. The error bars denote the 10th and 90th percentiles. All the models except the $+1\sigma$ and $-1\sigma$ have similar stellar mass distributions. The $+1\sigma$ and $-1\sigma$ models have corresponding shifts in their median stellar masses due to their selection.

\section{Discussion and Conclusions}
	
	We find that the Median, Mass Limit, Young Formation, Median Formation, Young Assembly and Median Assembly models have correlation lengths within the 68\% confidence interval for LAEs at $z=3.1$, all though the Young Formation and Young Assembly models have correlation lengths near the $+1\sigma$ limit. As designed, the $+\sigma$ and $-\sigma$ catalogs have $z=3.1$ correlation lengths near the observed $\pm 1\sigma$ correlation length uncertainty. The Young Merger and Median Merger models have large correlation lengths and are not consistent with the $z=3.1$ clustering measurement. We eliminated the Merger Median model due to the age of $\sim 1$ Gyr at $z=3.1$, in contrast with starburst ages of 20-100 Myr determined from SED fitting. However, SED modeling has not yet constrained the age of LAEs sufficiently to rule out the other age-based catalogs.
	
	We studied the connection between LAEs at $z=3.1$ and galaxy populations at $z=2.1$ for which previous studies have determined typical dark matter halo masses. \citet{Guaita:2010} found LAEs at $z=2.1$ to have a median dark matter halo mass of $\log(\textup{M/M}_\odot)=11.5^{+0.4}_{-0.5}$. Their study used bias evolution from the conditional mass function to show that $z=2.1$ LAEs are possible descendants of $z=3.1$ LAEs observed by Ga07. All of our models, except the $+\sigma$ and $-\sigma$ models, have median dark matter halo masses of the full descendant distributions around $10^{11.3}\textup{M}_\odot$ at $z=2.1$. Hence, $z=2.1$ LAEs could be direct descendants of $z=3.1$ LAEs, rather than a new set of halos undergoing their first phase of star formation. However, SED modeling by \citet{Guaita:2011} and \citet{Nilsson:2011} have found typical starburst ages for $z=2.1$ LAEs of $12^{+149}_{-3}\,\textup{Myr}$ and $80^{+10}_{-20}\,\textup{Myr}$. Given these starburst ages, the $z=2.1$ LAEs would have to be experiencing a subsequent burst of star formation to be descendants of $z=3.1$ LAEs. Another study by (\citealt{Adelberger:2005a}, see their Fig. 2) determined the clustering of BX galaxies at $z\simeq 2$. The clustering result for the least luminous ($K_s>21.5$) subset implies a median dark matter mass of $\log_{10}\textup{M}_{\textup{\tiny Med}}/\textup{M}_\odot=11.0^{+0.6}_{-0.9}$. This result is consistent with our $z=2.1$ descendant median dark matter mass, except for the $+\sigma$ and $-\sigma$ models. We find that $\sim75\%$ of the descendants of $z=3.1$ LAEs are centrals at $z=2.1$; it is unclear whether $\simeq2$ LAEs and BX galaxies represent a similar mix of centrals and satellites.
	
	We can make a similar comparison from $z=2.1$ to $z=0.9$. The Deep Extragalactic Evolutionary Probe 2 (DEEP2, \citealt{Davis:2003}) measured the clustering bias of color selected galaxies at $z\simeq 0.9$. For our chosen cosmology, their results \citep{Coil:2008} imply median dark matter halo masses of $\textup{M}_{\textup{\tiny Med}}=10^{12.9\pm0.1}\,\textup{M}_\odot$ and $\textup{M}_{\textup{\tiny Med}}=10^{12.0\pm0.1}\,\textup{M}_\odot$ for red and blue galaxies, respectively. The Mass limit, Median, and age-based models have $z=1$ descendant median dark matter halo masses of $\textup{M}\simeq 10^{11.9}\,\textup{M}_\odot$, consistent with the blue galaxies from the DEEP2 survey. The DEEP2 red galaxy subset are not consistent with being descendants of our LAE catalogs due to their larger dark matter mass. LAEs at $z=3.1$ are therefore possible progenitors of blue (late-type) galaxies residing in dark matter halos with mass $10^{12}\,\textup{M}_\odot$ at $z\simeq 1$. We find that $\sim65\%$ of the descendants of $z=3.1$ LAEs are centrals at $z=1$; it is unclear whether the blue DEEP2 galaxies represent a similar mix of centrals and satellites.
			
	The models that best represent LAEs at $z=3.1$ produce $z=0$ descendants with a median mass around $10^{12.7}\,\textup{M}_\odot$. We find that $\sim55\%$ of the descendants at $z=0$ are central sub-halos, with little variation on the fraction of the central descendants based on model selection. When we study only centrals, the age-based models have median dark matter halo masses at $z=0$ of $10^{12}\,\textup{M}_\odot$ while the Median Mass and Mass limit catalogs have slightly lower values of $10^{11.8}\,\textup{ M}_\odot$. We see no significant dependence on age definition or selection of the descendants' mass distributions. These results show that the Mass Limit, Median Mass, and age-based catalogs' descendants have a majority of central sub-halos which reside in $L^*$ type dark matter halos at $z=0$. However, the $+\sigma$ and $-\sigma$ models have descendant masses that are a factor of three larger and seven times smaller than the median mass of the Median model. The factor of five spread in median mass of the $+\sigma$ and $-\sigma$ catalogs' present-day descendants shows that a more precise measurement of the $z\simeq 3$ LAE bias is needed to place a stronger constraint on the descendant properties.

\acknowledgments
J. W. would like to thank Peter Kurczynski, Viviana Acquaviva, Caryl Gronwall, Lucia Guaita and Michael J. Berry for helpful comments of numerous drafts. This study was supported by grants from the National Science Foundation AST-0807570, 1055919 and Department of Energy DE-FG02-08ER41561. The Millennium Simulation databases used in this paper and the web application providing online access to them were constructed as part of the activities of the German Astrophysical Virtual Observatory. E.G. thanks the U. C. Davis Physics Department for hospitality during the completion of this research.


\clearpage

\clearpage


\begin{deluxetable}{lccccccccc}
	\tablecolumns{10}
	\tablewidth{0pc}
	\tablecaption{Mass Properties and correlation lengths for $z= 3.1$ LAE models}
	\tabletypesize{\tiny}
	\tablehead{\colhead{Name\tablenotemark{a}} & \colhead{25th \% Mass\tablenotemark{b}} & \colhead{Median Mass\tablenotemark{b}} & \colhead{75th \% Mass\tablenotemark{b}} & \colhead{$r_0$\tablenotemark{c}} & \colhead{$\gamma$\tablenotemark{c}} & \colhead{$\chi^2$/D.O.F.} & \colhead{$r_0$\tablenotemark{d}} & \colhead{$\gamma$\tablenotemark{d}} & \colhead{$\chi^2$/D.O.F.} \\
	\colhead{} & \colhead{$\log_{10}$(M$_\odot$)} & \colhead{$\log_{10}$(M$_\odot$)} & \colhead{$\log_{10}$(M$_\odot$)} & \colhead{Mpc} &  &  & \colhead{Mpc} &  &  }
	\startdata
	+1$\sigma$ & 11.36 & 11.40 & 11.44 & 4.67$^{+0.28}_{-0.25}$ & 1.80 & 66.53/18 & 4.37$^{+0.16}_{-0.15}$ & 1.34$^{+0.10}_{-0.09}$ & 9.29/17 \\
Mass Limit & 10.71 & 10.87 & 11.15 & 3.88$^{+0.30}_{-0.28}$ & 1.80 & 67.63/18 & 3.33$^{+0.17}_{-0.17}$ & 1.24$^{+0.10}_{-0.09}$ & 6.80/17 \\
Median Mass & 10.89 & 10.90 & 10.91 & 3.52$^{+0.21}_{-0.21}$ & 1.80 & 37.42/18 & 3.15$^{+0.16}_{-0.18}$ & 1.40$^{+0.12}_{-0.12}$ & 8.95/17 \\
-1$\sigma$ & 10.00 & 10.00 & 10.00 & 2.38$^{+0.25}_{-0.25}$ & 1.80 & 44.43/18 & 1.60$^{+0.32}_{-0.35}$ & 1.14$^{+0.19}_{-0.19}$ & 10.40/17 \\
\multicolumn{10}{l}{\textbf{Formation Age}} \\
Young Formation & 10.74 & 10.95 & 11.27 & 4.25$^{+0.28}_{-0.28}$ & 1.80 & 67.12/18 & 3.83$^{+0.17}_{-0.18}$ & 1.29$^{+0.11}_{-0.10}$ & 9.65/17 \\
Median Formation & 10.72 & 10.89 & 11.17 & 3.58$^{+0.25}_{-0.25}$ & 1.80 & 54.55/18 & 3.08$^{+0.17}_{-0.18}$ & 1.29$^{+0.11}_{-0.10}$ & 7.59/17 \\
\multicolumn{10}{l}{\textbf{Assembly Age}} \\
Young Assembly & 10.75 & 10.96 & 11.28 & 4.37$^{+0.32}_{-0.30}$ & 1.80 & 84.68/18 & 3.90$^{+0.20}_{-0.21}$ & 1.25$^{+0.12}_{-0.12}$ & 13.05/17 \\
Median Assembly & 10.73 & 10.90 & 11.18 & 3.56$^{+0.18}_{-0.21}$ & 1.80 & 35.41/18 & 3.19$^{+0.17}_{-0.18}$ & 1.41$^{+0.11}_{-0.11}$ & 8.41/17 \\
\multicolumn{10}{l}{\textbf{Merger Age}} \\
Young Merger & 10.72 & 10.91 & 11.19 & 4.65$^{+0.25}_{-0.28}$ & 1.80 & 62.75/18 & 4.31$^{+0.14}_{-0.14}$ & 1.33$^{+0.09}_{-0.09}$ & 7.42/17 \\
Merger Median & 10.71 & 10.87 & 11.15 & 5.04$^{+0.39}_{-0.37}$ & 1.80 & 121.25/18 & 4.66$^{+0.16}_{-0.16}$ & 1.17$^{+0.09}_{-0.09}$ & 8.61/17
	\enddata
	\tablenotetext{a}{All models have a number density of $1.5\times 10^{-3}\,\textup{Mpc}^{-3}$. $+1\sigma$: Centered around $\log_{10}\textup{M}=11.4\textup{M}_\odot$; Mass Limit: Halos with M$\ge 3.98\times 10^{10}\,\textup{M}_\odot$; Median Mass: Centered around log$_{10}$M=10.9 M$_{\odot}$; $-1\sigma$: Centered around $\log_{10}\textup{M}=10.0\,\textup{M}_{\odot}$; Young Formation: Youngest sub-halos based on Formation age; Median Formation: Centered around the median 0.744 Gyr.; Young Assembly: Youngest sub-halos based on Assembly age; Median Assembly:  Centered around the median 0.943 Gyr.; Young Merger: Youngest sub-halos based on Merger age.; Median Merger: Centered around the median 1.090 Gyr.}
	\tablenotetext{b}{Mass reported is the mass of the FOF halo where the sub-halo resides, $\textup{M}_{\textup{FOF}}$.}
	\tablenotetext{c}{Parameters for fits using fixed $\gamma=1.8$.}
	\tablenotetext{d}{Parameters for fits allowing both $r_0$ and $\gamma$ as free parameters.}	
	\label{model:mass}
\end{deluxetable}


\setlength{\tabcolsep}{0.024in}

\begin{deluxetable}{lccccccccc}

	\tablecolumns{10}
	\tabletypesize{\scriptsize}
	\tablewidth{\linewidth}
	\tablecaption{Age Properties of $z=3.1$ LAE models}
	\tablehead{\colhead{Name\tablenotemark{a}} & \colhead{Form. 25th} & \colhead{Form. 50th} & \colhead{Form. 75th} & \colhead{Assem. 25th} & \colhead{Assem. 50th} & \colhead{Assem. 75th} & \colhead{Merg. 25th} & \colhead{Merg. 50th} & \colhead{Merg. 75th} \\
	& \colhead{Gyr} & \colhead{Gyr} & \colhead{Gyr} & \colhead{Gyr} & \colhead{Gyr} & \colhead{Gyr} & \colhead{Gyr} & \colhead{Gyr} & \colhead{Gyr}}
	\startdata
	+1$\sigma$ & 0.523 & 0.696 & 0.857 & 0.644 & 0.864 & 1.033 & 0.583 & 0.978 & 1.454 \\
Mass Limit & 0.569 & 0.748 & 0.921 & 0.724 & 0.946 & 1.110 & 0.583 & 1.090 & 1.527 \\
Median Mass & 0.580 & 0.766 & 0.942 & 0.754 & 0.964 & 1.128 & 0.583 & 1.090 & 1.527 \\
-1$\sigma$ & 0.675 & 0.877 & 1.066 & 0.915 & 1.080 & 1.218 & 0.725 & 1.288 & 1.802 \\
\multicolumn{10}{l}{\textbf{Formation Age}} \\
Young Formation & \textbf{0.156} & \textbf{0.258} & \textbf{0.301} & 0.164 & 0.289 & 0.354 & 0.267 & 0.856 & 1.454 \\
Median Formation & \textbf{0.738} & \textbf{0.747} & \textbf{0.755} & 0.831 & 0.905 & 1.018 & 0.583 & 0.978 & 1.454 \\
\multicolumn{10}{l}{\textbf{Assembly Age}} \\
Young Assembly & 0.158 & 0.270 & 0.324 & \textbf{0.164} & \textbf{0.289} & \textbf{0.345} & 0.583 & 1.090 & 1.593 \\
Median Assembly & 0.684 & 0.785 & 0.847 & \textbf{0.936} & \textbf{0.946} & \textbf{0.956} & 0.583 & 0.978 & 1.454 \\
\multicolumn{10}{l}{\textbf{Merger Age}} \\
Young Merger & 0.419 & 0.553 & 0.686 & 0.693 & 0.897 & 1.069 & \textbf{0.091} & \textbf{0.091} & \textbf{0.091} \\
Merger Median & 0.640 & 0.832 & 0.981 & 0.749 & 0.970 & 1.120 & \textbf{0.978} & \textbf{1.090} & \textbf{1.090}
	\enddata

	\tablenotetext{a}{Model descriptions found in Table \ref{model:mass}.}
	\tablecomments{Table reports the 25th, 50th and 75th percentiles for each model using the Formation (Form.), Assembly (Assem.) and Merger (Merg.) age definitions (See section \ref{agecat}). Bold entries mark the values in the age definition used to select the age-based models.} 
	\label{model:age}
\end{deluxetable}

\begin{deluxetable}{lccccc}
	\tablecolumns{6}
	\tablewidth{0pc}
	\tablecaption{$z=2.1$ Descendant Mass Distributions}
	\tabletypesize{\tiny}
	\tablehead{\colhead{Name\tablenotemark{a}} & \colhead{Distribution Type\tablenotemark{b}} & \colhead{Fraction\tablenotemark{c}} & \colhead{25th \% Mass\tablenotemark{d}} & \colhead{Median Mass\tablenotemark{d}} & \colhead{75th \% Mass\tablenotemark{d}} \\
	\colhead{} & \colhead{} & \colhead{} & \colhead{$\log_{10}$(M$_\odot$)} & \colhead{$\log_{10}$(M$_\odot$)} & \colhead{$\log_{10}$(M$_\odot$)}}
	\startdata
	\textbf{+1$\sigma$} & {\it Full} & 100\% & 11.54 & 11.66 & 11.94 \\
& {\it Central} & 76\% & 11.51 & 11.60 & 11.72 \\
& {\it Satellite} & 24\% & 12.05 & 12.33 & 12.80 \\
\textbf{Mass Limit} & {\it Full} & 100\% & 10.94 & 11.25 & 11.77 \\
& {\it Central} & 76\% & 10.88 & 11.10 & 11.42 \\
& {\it Satellite} & 24\% & 11.64 & 12.06 & 12.57 \\
\textbf{Median Mass} & {\it Full} & 100\% & 11.01 & 11.13 & 11.50 \\
& {\it Central} & 73\% & 10.99 & 11.06 & 11.18 \\
& {\it Satellite} & 27\% & 11.58 & 11.92 & 12.41 \\
\textbf{-1$\sigma$} & {\it Full} & 100\% & 10.09 & 10.20 & 10.65 \\
& {\it Central} & 72\% & 10.06 & 10.13 & 10.24 \\
& {\it Satellite} & 28\% & 10.76 & 11.24 & 11.95 \\
\textbf{Young Formation} & {\it Full} & 100\% & 11.01 & 11.36 & 11.88 \\
& {\it Central} & 72\% & 10.93 & 11.18 & 11.59 \\
& {\it Satellite} & 28\% & 11.50 & 11.91 & 12.51 \\
\textbf{Median Formation} & {\it Full} & 100\% & 10.94 & 11.25 & 11.74 \\
& {\it Central} & 77\% & 10.88 & 11.11 & 11.43 \\
& {\it Satellite} & 23\% & 11.65 & 12.06 & 12.52 \\
\textbf{Young Assembly} & {\it Full} & 100\% & 11.01 & 11.36 & 11.91 \\
& {\it Central} & 72\% & 10.92 & 11.19 & 11.58 \\
& {\it Satellite} & 28\% & 11.52 & 11.94 & 12.55 \\
\textbf{Median Assembly} & {\it Full} & 100\% & 10.97 & 11.26 & 11.71 \\
& {\it Central} & 77\% & 10.91 & 11.12 & 11.44 \\
& {\it Satellite} & 23\% & 11.57 & 11.97 & 12.47 \\
\textbf{Young Merger} & {\it Full} & 100\% & 10.96 & 11.30 & 11.79 \\
& {\it Central} & 75\% & 10.90 & 11.12 & 11.48 \\
& {\it Satellite} & 25\% & 11.64 & 12.04 & 12.56 \\
\textbf{Merger Median} & {\it Full} & 100\% & 10.94 & 11.24 & 11.76 \\
& {\it Central} & 76\% & 10.89 & 11.10 & 11.40 \\
& {\it Satellite} & 24\% & 11.65 & 12.06 & 12.58
	\enddata
	\tablenotetext{a}{Model descriptions found in Table \ref{model:mass}.}
	\tablenotetext{b}{Centrals are the most massive sub-halos within their FOF halo. Satellites are all other sub-halos. Full describes the properties of all descendant sub-halos. The median mass values for the central and satellite distributions are shown in Figures \ref{massplot} and \ref{massplotage}.}
	\tablenotetext{c}{Fraction of objects within full, central or satellite subset.}
	\tablenotetext{d}{Mass reported is the mass of the FOF halo where the sub-halo resides, $\textup{M}_{\textup{FOF}}$.}
	\label{model:massdesc2}
\end{deluxetable}

\begin{deluxetable}{lccccc}
	\tablecolumns{6}
	\tablewidth{0pc}
	\tablecaption{$z=1$ Descendant Mass Distributions}
	\tabletypesize{\tiny}
	\tablehead{\colhead{Name\tablenotemark{a}} & \colhead{Distribution Type\tablenotemark{b}} & \colhead{Fraction\tablenotemark{c}} & \colhead{25th \% Mass\tablenotemark{d}} & \colhead{Median Mass\tablenotemark{d}} & \colhead{75th \% Mass\tablenotemark{d}} \\
	\colhead{} & \colhead{} & \colhead{} & \colhead{$\log_{10}$(M$_\odot$)} & \colhead{$\log_{10}$(M$_\odot$)} & \colhead{$\log_{10}$(M$_\odot$)}}
	\startdata
	\textbf{+1$\sigma$} & {\it Full} & 100\% & 11.83 & 12.21 & 12.90 \\
& {\it Central} & 63\% & 11.74 & 11.92 & 12.20 \\
& {\it Satellite} & 37\% & 12.70 & 13.10 & 13.54 \\
\textbf{Mass Limit} & {\it Full} & 100\% & 11.30 & 11.86 & 12.68 \\
& {\it Central} & 62\% & 11.12 & 11.43 & 11.88 \\
& {\it Satellite} & 38\% & 12.23 & 12.81 & 13.42 \\
\textbf{Median Mass} & {\it Full} & 100\% & 11.27 & 11.68 & 12.53 \\
& {\it Central} & 61\% & 11.18 & 11.34 & 11.63 \\
& {\it Satellite} & 39\% & 12.21 & 12.74 & 13.33 \\
\textbf{-1$\sigma$} & {\it Full} & 100\% & 10.29 & 10.70 & 11.93 \\
& {\it Central} & 61\% & 10.21 & 10.34 & 10.61 \\
& {\it Satellite} & 39\% & 11.43 & 12.18 & 13.04 \\
\textbf{Young Formation} & {\it Full} & 100\% & 11.40 & 12.00 & 12.79 \\
& {\it Central} & 60\% & 11.17 & 11.54 & 12.05 \\
& {\it Satellite} & 40\% & 12.27 & 12.80 & 13.44 \\
\textbf{Median Formation} & {\it Full} & 100\% & 11.30 & 11.84 & 12.63 \\
& {\it Central} & 63\% & 11.13 & 11.45 & 11.89 \\
& {\it Satellite} & 37\% & 12.21 & 12.78 & 13.38 \\
\textbf{Young Assembly} & {\it Full} & 100\% & 11.40 & 12.00 & 12.79 \\
& {\it Central} & 60\% & 11.18 & 11.55 & 12.06 \\
& {\it Satellite} & 40\% & 12.26 & 12.79 & 13.44 \\
\textbf{Median Assembly} & {\it Full} & 100\% & 11.32 & 11.83 & 12.61 \\
& {\it Central} & 63\% & 11.15 & 11.46 & 11.87 \\
& {\it Satellite} & 37\% & 12.19 & 12.71 & 13.38 \\
\textbf{Young Merger} & {\it Full} & 100\% & 11.34 & 11.91 & 12.67 \\
& {\it Central} & 62\% & 11.16 & 11.49 & 11.97 \\
& {\it Satellite} & 38\% & 12.19 & 12.75 & 13.39 \\
\textbf{Merger Median} & {\it Full} & 100\% & 11.27 & 11.83 & 12.61 \\
& {\it Central} & 63\% & 11.12 & 11.42 & 11.87 \\
& {\it Satellite} & 37\% & 12.23 & 12.78 & 13.45
	\enddata
	\label{model:massdesc1}
	\tablenotetext{a}{Model descriptions found in Table \ref{model:mass}.}
	\tablenotetext{b}{As defined in Table 3. The median mass values for the central and satellite distributions are shown in Figures \ref{massplot} and \ref{massplotage}.}
	\tablenotetext{c}{Fraction of objects within full, central or satellite subset.}
	\tablenotetext{d}{Mass reported is the mass of the FOF halo where the sub-halo resides, $\textup{M}_{\textup{FOF}}$.}
\end{deluxetable}

\begin{deluxetable}{lccccc}
	\tablecolumns{6}
	\tablewidth{0pc}
	\tablecaption{$z=0$ Descendant Mass Distributions}
	\tabletypesize{\tiny}
	\tablehead{\colhead{Name\tablenotemark{a}} & \colhead{Distribution Type\tablenotemark{b}} & \colhead{Fraction\tablenotemark{c}} & \colhead{25th \% Mass\tablenotemark{d}} & \colhead{Median Mass\tablenotemark{d}} & \colhead{75th \% Mass\tablenotemark{d}} \\
	\colhead{} & \colhead{} & \colhead{} & \colhead{$\log_{10}$(M$_\odot$)} & \colhead{$\log_{10}$(M$_\odot$)} & \colhead{$\log_{10}$(M$_\odot$)}}
	\startdata
	\textbf{+1$\sigma$} & {\it Full} & 100\% & 12.26 & 12.95 & 13.83 \\
& {\it Central} & 60\% & 12.05 & 12.38 & 12.93 \\
& {\it Satellite} & 40\% & 13.41 & 13.88 & 14.36 \\
\textbf{Mass Limit} & {\it Full} & 100\% & 11.75 & 12.65 & 13.69 \\
& {\it Central} & 58\% & 11.42 & 11.93 & 12.61 \\
& {\it Satellite} & 42\% & 13.04 & 13.70 & 14.32 \\
\textbf{Median Mass} & {\it Full} & 100\% & 11.68 & 12.51 & 13.57 \\
& {\it Central} & 55\% & 11.44 & 11.77 & 12.32 \\
& {\it Satellite} & 45\% & 12.97 & 13.60 & 14.13 \\
\textbf{-1$\sigma$} & {\it Full} & 100\% & 10.63 & 11.71 & 13.26 \\
& {\it Central} & 57\% & 10.42 & 10.73 & 11.50 \\
& {\it Satellite} & 43\% & 12.29 & 13.24 & 13.96 \\
\textbf{Young Formation} & {\it Full} & 100\% & 11.91 & 12.84 & 13.74 \\
& {\it Central} & 57\% & 11.53 & 12.07 & 12.86 \\
& {\it Satellite} & 43\% & 13.08 & 13.70 & 14.23 \\
\textbf{Median Formation} & {\it Full} & 100\% & 11.79 & 12.64 & 13.66 \\
& {\it Central} & 56\% & 11.45 & 11.94 & 12.55 \\
& {\it Satellite} & 44\% & 12.99 & 13.63 & 14.22 \\
\textbf{Young Assembly} & {\it Full} & 100\% & 11.92 & 12.85 & 13.76 \\
& {\it Central} & 57\% & 11.56 & 12.08 & 12.86 \\
& {\it Satellite} & 43\% & 13.09 & 13.72 & 14.32 \\
\textbf{Median Assembly} & {\it Full} & 100\% & 11.79 & 12.60 & 13.58 \\
& {\it Central} & 58\% & 11.47 & 11.92 & 12.52 \\
& {\it Satellite} & 42\% & 13.00 & 13.63 & 14.20 \\
\textbf{Young Merger} & {\it Full} & 100\% & 11.83 & 12.71 & 13.70 \\
& {\it Central} & 58\% & 11.47 & 11.99 & 12.67 \\
& {\it Satellite} & 42\% & 13.03 & 13.71 & 14.20 \\
\textbf{Merger Median} & {\it Full} & 100\% & 11.75 & 12.56 & 13.73 \\
& {\it Central} & 58\% & 11.42 & 11.88 & 12.51 \\
& {\it Satellite} & 42\% & 12.96 & 13.77 & 14.37
	\enddata
	\tablenotetext{a}{Model descriptions found in Table \ref{model:mass}.}
	\tablenotetext{b}{As defined in Table 3. The median mass values for the central and satellite distributions are shown in Figures \ref{massplot} and \ref{massplotage}.}
	\tablenotetext{c}{Fraction of objects within full, central or satellite subset.}
	\tablenotetext{d}{Mass reported is the mass of the FOF halo where the sub-halo resides, $\textup{M}_{\textup{FOF}}$.}
\label{model:massdesc0}
\end{deluxetable}

\end{document}